\documentclass[10pt, conference, compsocconf]{IEEEtran}

\usepackage{amsmath}
\usepackage{listings } \usepackage{url}
\usepackage{graphicx}

\renewcommand\footnotemark{}

\begin{document}

\title{Improving software team collaboration with Synchronized
Software Development}
\date{October 31, 475}
\author{\IEEEauthorblockN{Stanislav Levin}
\IEEEauthorblockA{The Blavatnik School of Computer Science\\
 Tel Aviv University\\
Tel-Aviv, Israel\\
stanisl@post.tau.ac.il}
\and
\IEEEauthorblockN{Amiram Yehudai}
\IEEEauthorblockA{The Blavatnik School of Computer Science\\
 Tel Aviv University\\
Tel-Aviv, Israel\\
amiramy@tau.ac.il}
}

\maketitle

\let\thefootnote\relax\footnote{This research was supported by THE ISRAEL SCIENCE FOUNDATION, grant No. 476/11. This paper was written in 2012.}

\begin{abstract}

Effective collaboration is a key factor in the success of a software project
developed by a team. In this work, we suggest the approach of Synchronized
Software Development (SSD), which promotes a new mechanism of collaboration in
general, and for code synchronization in particular. In SSD, code changes made
by one developer are automatically propagated to others as long as they keep the
code free of compilation errors. Changes that introduce compilation errors are
not propagated until the errors are fixed. Moreover, other developers are
restricted from concurrently editing the entities involved in these changes.
While in this state, developers are, however, free to modify the rest of the
entities.

The novelty of our approach is that it actively synchronizes developers with the
latest error free version of the source code, preventing possible conflicts and
merges that may arise due to concurrent changes made by fellow team members. SSD
also allows for a more transparent an practically near real time awareness of
new code that is being introduced by multiple developers. We built CSI (Code
Synchronizing Intelligence), a prototype demonstrating key features of SSD. 
\end{abstract}

\begin{IEEEkeywords}
collaboration; change; awareness; conflicts; merge; 
\end{IEEEkeywords}

\section{Introduction}

Modern software projects involve multiple developers collaboratively working on
the same codebase. In fact, parallel development has become the norm rather an
exception \cite{SupportingIndirectConflicts}. The task of sharing a codebase
repository is usually carried out by a Software Configuration Management system
(SCM) \cite{SVN, ClearCase, GIT, Mercurial}. The SCM system maintains all files
that comprise the software project, and serves as the only version controlling
mechanism through which developers share code \cite{EnvironmentForSyncDev}. The
SCM tools employ a common checkin / checkout model according to which a change
will become visible to others, only after the developer who made it checks in
his code to the shared repository. A direct implication of this model is that
code conflicts will only be discovered post factum, when a developer tries to
checkin the already conflicting code. Once aware of the conflict the developer
is forced to resolve it by means of merging his version with the repository's
one. Such manual merges are considered both time consuming and error prone
\cite{ConcurrencyControlInGroupwareSystems, ProactiveConflictDetection}. This is
a definite limitation of the current checkin / checkout model. Inspired by
Google Docs \cite{GoogleDocs}, we envision a development environment that
provides similar real time collaborative editing capabilities for code
development.

\section{Related work}
A number of tools have been suggested so as to address the code conflicts issue,
and have mostly concentrated on increasing developers' awareness of the
activities performed by fellow team members. Such tools usually come either as a
plug-in integrated into the Integrated Development Environment (IDE) (Syde
\cite{Syde}, Lighthouse \cite{Lighthouse}) or as a standalone application
(Palant\'ir \cite{Palantir}, FASTDash \cite{FASTDash}). CollabVS \cite{CollabVS}
extends the standard Visual Studio user interface with collaboration oriented
features such as chat, video, and audio streams on top of its conflict detection
mechanisms. CloudStudio \cite{CloudStudio} suggests a concept of "cloud-based
development", and replaces the explicit checkin / checkout model with
interactive editing and real-time conflict tracking and management. One of the
main goals of these tools is to provide developers with relevant information so
as to assist them in avoiding code conflicts. The described tools share a common
principle, they collect relevant information and present it to the developer. It
is up to the developer to utilize this information and perform (or refrain from
performing) a particular set of actions in order to prevent conflicts and
promote collaboration. The developer is only shown the path, yet he is the one
who has to walk it.

Our work addresses the difference between knowing the path and walking the path
\cite{Matrix}. The suggested approach, Synchronized Software Development (SSD),
forcibly turns concurrent changes into sequential ones, by allowing only one
developer to edit any given entity (e.g. method) at any given time. Other
developers are blocked from concurrently editing that particular entity. While
blocked they may, however, edit other entities in the code. In addition to the
trivial case of concurrently editing the same entity, SSD also aims to detect
indirectly conflicting (concurrent) changes that may affect one another, 
resulting in a conflict. For instance, changing a method's name at one site,
while using the old method's name at another. SSD strives to enforce conflict
prevention by means of fine grained restrictions on entity editing, and thus
prevent the conflicts (and \textit{manual} merges) originating from the checkin
/ checkout synchronization model typically employed by current SCM systems.

\section{Motivating Use Case}

We call the state in which the code fails to compile \emph{unbuildable} state,
and the state in which the code successfully compiles a \emph{buildable} state.

We shall now analyze a use case demonstrating SSD in a real life scenario, 
accompanied with some screenshots from CSI, our SSD prototype.

\begin{enumerate}

\item 
Two developers, Alice and Bob begin in the same file state.
See figure \ref{fig:UseCaseStep1}, section 1.

\begin{figure*}
\caption{\newline (1) Alice and Bob begin in the same state. \newline(2) Bob
introduces some changes. Method's name is changed, and the new parameter has an
invalid type.}
\label{fig:UseCaseStep1}
\includegraphics[scale=0.62]{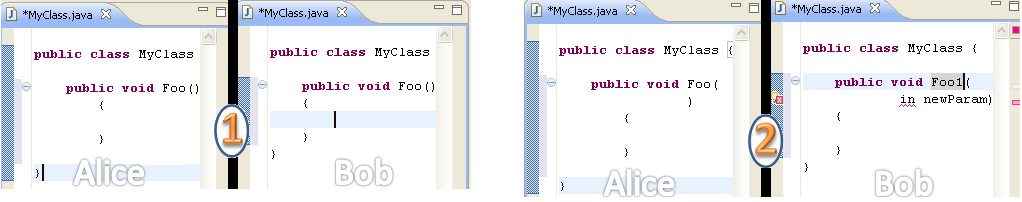}
\end{figure*}

\item
Bob intends to add a new parameter, "newParam", of type int to the
method "Foo". Bob begins typing in his change, but mistakenly types "in
newParam", missing the "t" at the end of "int".

\item 
Not being aware of his mistyping, Bob also changes the name of the method from
"Foo" to "Foo1". At this stage Bob has a new name for the method "Foo" (i.e.
"Foo1") and an additional, incorrect parameter definition. These changes render
the file state unbuildable. Since the code state is currently unbuildable, none
of the changes Bob has made are propagated to Alice. See figure
\ref{fig:UseCaseStep1}, section 2.

\item
Meanwhile, Alice intends to change the method "Foo" to "Foo2". She is currently
unaware that Bob has already changed this method's name to "Foo1", and in the
file version she currently has, the method's name is still "Foo". Alice begins
changing the name of "Foo" to "Foo2", say by typing an additional "2" at the end
of the "Foo" string in the method definition, and is immediately warned that the
current method is locked for editing by another developer (Bob).

 \item
Alice is now \textit{made aware} that the method is undergoing changes by some
other developer (Bob), and is forced to wait till these changes are complete,
avoiding the conflict that would have otherwise been introduced due to fact they
both had changed the method's name. See figure \ref{fig:UseCaseStep3}, section
3.

\begin{figure*}
\center
\caption{\newline(3) Alice's trial to edit the method's name prompts the system
to block her change, and display an alert.\newline(4) Bob fixes the errors and
turns the file state buildable, his code is then propagated to Alice.}
\label{fig:UseCaseStep3}
\includegraphics[scale=0.6]{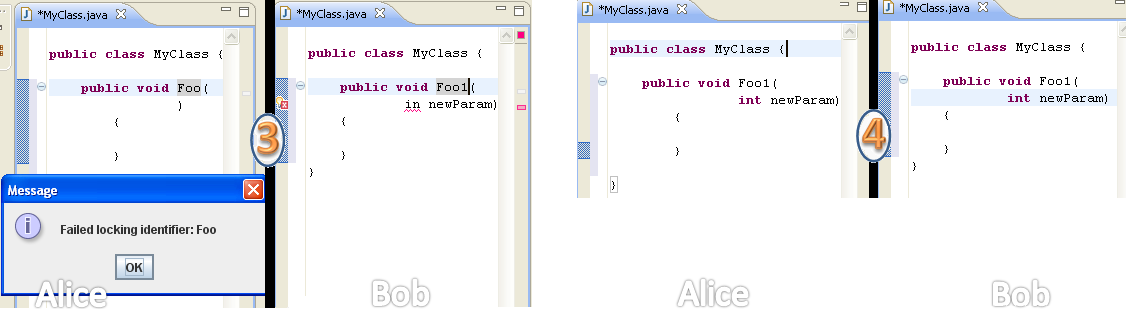}
\end{figure*}

 \item 
Once Bob adds the "t" to the mistyped "in", his code becomes buildable, and is
instantly propagated to Alice, which gets the new method name, with the
additional parameter added by Bob. Alice may now commence the change she has
intended. See figure \ref{fig:UseCaseStep3}, section 4.


\end{enumerate}

It is worth noting that SSD works on the fly, as developers type in code.
Neither Alice nor Bob has to actively save their file in order for SSD to
perform. This may be witnessed by the asterisk symbol near the file name at the
top of the editing tab in the Eclipse IDE, which indicates that the file at hand
has not been saved yet and all changes are currently buffered in memory, see
figures \ref{fig:UseCaseStep1}, \ref{fig:UseCaseStep3}.

In a typical SCM system, such a conflict would only be discovered post factum,
when a developer would try to checkin an already conflicting code version. In an
SSD system the conflict is discovered \textbf{at the time of producing conflicting
code}, while in current SCM systems it is only discovered
\textbf{after it has been checked in to the SCM}.

The key concept of an SSD system is that it is aware of all changes currently
carried by all team members. Thus, once the (chronologically) first developer
begins changing the method's name, the SSD system locks the method element for
editing by other developers. When another developer tries to change the same
method, the SSD system will notify him that the element he's trying to edit is
already being edited. He should then wait until the undergoing change is
complete, and only then introduce his own changes. By means of locking we aim at
preventing concurrent, conflicting changes, that otherwise might have resulted
in a conflict. However, the locking is so fine grained that we expect it to be
practically transparent to developers, and they will only be aware of it in
case it intervenes to prevent a highly probable conflict.

The concept demonstrated in the use case we described applies to a wide variety
of changes: introducing new methods, changing existing method's name, changing
existing method's body, and so on. An SSD system should support all code
editing operations available in a standard IDE.

\section{Dependency detection to the aid of conflicts prevention}

We believe it is highly undesirable for developers to make design related
decisions based on stale code. Our fundamental assumption is that while
coding, a developer would rather wait (obviously, within reason), than engage in
a manual merge process incurred by possible code conflicts. Our efforts are
proactive, directed at preventing conflicts before they actually occur.

We establish the notion of element (i.e., Abstract Syntax Tree nodes (AST))
dependency. Elements $E_1$, $E_2$ are {\em dependent} if one of the following
holds:

\begin{enumerate}
 \item $E_1 = E_2$.
 \item $E_1, E2$ have a common ancestor of type method or statement in the AST .
\item $E_1$ references $E_2$'s binding or vice versa (e.g., $E_1=aMemberField + 1;$
 $E_2= int\;aMemberField;$).
\end{enumerate}

We argue that in order to prevent conflicts, no dependent elements
should be subject to concurrent editing.

The first case implies that no single element may be concurrently edited.

The second case deals with concurrent editing of elements such as statements
inside methods, or parameters in method invocations taking place inside the body
block of another method. Theoretically, an SSD system could allow concurrent
editing of statements in the same method, however, we believe this is not a good
practice since it may lead to inconsistencies in the method's logic.

We demonstrate the third case with an example. Suppose we have a variable:
$$int\; someVar;$$ denoted by $E_1$, and a statement: $$int\; otherVar =
someVar;$$ denoted by $E_2$. Let $E_{i_{d_j}}$ denote $E_i$ element's copy at
developer $j$'s site (i.e., $E_{1_{d_2}}$ is $E_1$'s copy at developer2's site).

Suppose now, that developer1 renames $E_{1_{d_1}}$ to: $$int\; newSomeVar;$$
while at the same time, developer2 changes $E_{2_{d_2}}$ to: $$int\; otherVar =
someVar + 1;$$ (before either change is propagated to the other site). Note that
developer1's renaming of $E_{1_{d_1}}$ results in a cascading change to
$E_{2_{d_1}}$ in order to make the code buildable. $E_{2_{d_1}}$ is now: $$int\;
otherVar = newSomeVar ;$$ and it is in conflict with $E_{2_{d_2}}$, which is:
$$int\; otherVar = someVar + 1;$$ Once such a state is reached, no matter what
order the changes are propagated in, a conflict is inevitable, $E_{2_{d_2}} \neq
E_{2_{d_1}}$. The third element dependency condition above aims to prevent such
cases.

\section{CSI - An SSD prototype}

We've begun implementing CSI (Code Synchronizing Intelligence), an SSD
prototype plug in for the Eclipse IDE. CSI uses the Java Model
\cite{JavaModel} offered by the Eclipse JDT (Java Development tools) in order to
be notified of changes (introduction, deletion and modification of Java
elements, which in turn may be classes, methods, member variables and so on)
made to the model representing the program structure. The Java Model plays an
important role in tracking changes on a semantical level, rather than observing
textual changes, which is a key principle of SSD.

\section{Discussion \& Limitations}

It is crucial to determine the dependent elements as soon as possible in order
to enforce locking in near real time. Any delay in doing so may result in a
conflict due to unrestricted concurrent editing. We're investigating ways to
employ AST resolution on the fly (as code is written) in order to detect complex
element dependency, while incurring minimal performance cost.

Developer's privacy should also be taken into account. A developer may want to
go "off the record" whenever he wishes to delay the propagation of his changes,
despite the fact that technically, they can be propagated immediately. Local
testing is great motivation for going off record. However, this
increases the chance of introducing a conflict once going back "on record", since
during the off record period the developer is unsupervised by the locking mechanism.

\section{Impact and Future Work\label{conc}}

SSD combined with cloud computing in general, and cloud-based development
\cite{CloudStudio} in particular, will  result in a powerful, collaboration
oriented IDE, provided in a form of Software as a Service (SaaS) \cite{SaaS}.
This, in turn, may change the current perception of an IDE, and present
opportunities for new paradigms in the field of software development.

An example of such a paradigm is near real time automated unit testing. Near
real time code propagation presents the opportunity for running automated unit
test suites on code that has just been written (long before it is checked-in to the
SCM). To reduce running times, regression test selection algorithms and techniques
\cite{RegressionTestSelection} can be employed. Near real time
regression testing will assist in a considerably earlier regression bug
detection, than in a traditional checkin / checkout model. This in turn, will
lead to a cost reduction in software development projects
\cite{HowTestsDriveTheCode}.

SSD can change the rules of known practices like Pair Programming, challenging
the traditional separation between the driver-navigator roles.

The nature of SSD blurs the boundaries between distributed and non
distributed software development,  enabling close collaboration even between
geographically separated developers. 

We intend to elaborate our research and extend our vision of software
development environments in the presence of SSD. Our future efforts will be
dedicated to conducting further user studies and experiments in order to devise
SSD best practices.

\bibliographystyle{IEEEtran}
\bibliography{SSD}

\end{document}